\documentclass[prb,aps,10pt,twocolumn,showpacs,floats,amsmath,amssymb,groupedaddress]{revtex4-1}

\usepackage{hyperref}
\usepackage{graphicx}% Include figure files
\usepackage{dcolumn}% Align table columns on decimal point
\usepackage{bm}% bold math
\usepackage{amssymb}% for unusual symbols
\usepackage[latin1]{inputenc}
\usepackage{subfigure}
\usepackage{color}
\usepackage{xcolor}
\usepackage{pifont,wasysym}
\usepackage{enumerate}
\usepackage{ulem}   % pour barrer du texte

\def \be{\begin{equation}}
\def \ee{\end{equation}}

\definecolor{orange}{rgb}{1,0.5,0}
\definecolor{darkgreen}{rgb}{0.13, 0.55, 0.13}

\newcommand\commentout[1]{}

\begin{document}

\title{Graphene \textit{n-p} junctions in the quantum Hall regime:\\
numerical study of incoherent scattering effects}

\author{Qianfan Ma}
\author{Fran\c{c}ois D. Parmentier}
\author{Preden Roulleau}
\author{Genevi\`eve Fleury}
\email{genevieve.fleury@cea.fr}
\affiliation{SPEC, CEA, CNRS, Universit\'e Paris-Saclay,\\
CEA Saclay, 91191 Gif-sur-Yvette Cedex, France}

\begin{abstract}
We investigate electronic transport through a graphene \textit{n-p} junction in the quantum Hall effect regime at high perpendicular magnetic field, when the filling factors in the \textit{n}-doped and \textit{p}-doped regions are fixed to 2 and -2 respectively. We compute numerically the conductance $G$, the noise $Q$ and the Fano factor $F$ of the junction when inelastic effects are included along the interface in a phenomenological way, by means of fictitious voltage probes. 
Using a scaling approach, we extract the system coherence length $L_\phi$ and describe the full crossover between the coherent limit ($W\ll L_\phi$) and the incoherent limit ($W\gg L_\phi$), $W$ being the interface length. While $G$ saturates at the value $e^2/h$ in the incoherent regime, $Q$ and $F$ are found to vanish exponentially for large length $W$. Corrections due to disorder are also investigated. Our results are finally compared to available experimental data.
\end{abstract}


\maketitle
\section{Introduction}

\commentout{electron focusing\cite{Chen2016}
In the route towards graphene-based electron optics, the graphene\textit{n-p} junction is a fascinating playground. The interface between the \comment{\textit{n-}} and \comment{\textit{p-}}doped regions can be used to guide carriers along snake states over several microns~\cite{Williams2011bis,Rickhaus2015bis,Taychatanapat2015} while a parabolic-shaped \textit{n-p} junction can be used as a lens to create robust collimated electron beams~\cite{Liu2017}. Alternative electron guiding has been demonstrated in Refs.\cite{Williams2011,Rickhaus2015} using gate-defined pn junctions. Fabry Perot interference patterns \cite{Rickhaus2013,Grushina2013} as well as magnetoresistance oscillations\cite{Overweg2017} have been observed in graphene pn junctions.} 
\commentout{In the route towards graphene-based electron optics, the graphene \textit{n-p} junction is a fascinating playground.\cite{Rickhaus2013,Grushina2013,Chen2016,Overweg2017}}

Graphene \textit{n-p} junctions are a fascinating playground for the implementation of electron optics experiments.\cite{Rickhaus2013,Grushina2013,Chen2016,Overweg2017,Jiang2017}
The possibility to guide charge carriers over several microns, using snake states,\cite{Williams2011bis,Milovanovic2014,Rickhaus2015bis,Taychatanapat2015} gate-defined electron waveguides,\cite{Williams2011,Rickhaus2015} or lensing apparatus\cite{Liu2017} has attracted growing interest.
In the quantum Hall regime, chiral edge states provide natural electron beams. 
Electron- and hole-like edge channels, which propagate in opposite directions in the \textit{n-} and \textit{p-} regions respectively, meet at the \textit{n-p} junction and co-propagate along its interface. The co-propagating channels then split toward their respective regions upon reaching the end of the junction.
\commentout{Electron- and hole-like edge channels counter-propagate in the n and p parts, meet at the \textit{n-p} junction, co-propagate along the interface and finally split when they reach the other end of the junction.} 
In addition to conductance measurements,\cite{Williams2007,Ozyilmaz2007,Ki2009,Lohmann2009,Velasco2009,Ki2010,Woszczyna2011,Schmidt2013,Amet2014,Klimov2015,Matsuo2015bis,Tovari2016} shot noise measurements\cite{Matsuo2015,Kumada2015} have demonstrated that a graphene \textit{n-p} junction can act as a coherent beam splitter of electron-like and hole-like particles. Recently, a Mach-Zehnder edge-channel interferometer has been implemented in such a device, showing robust conductance oscillations with very high visibility.\cite{Wei2017} \\
\indent When the filling factors in the \textit{n-} and \textit{p-} parts are tuned to $\nu_n=2$ and $\nu_p=-2$ respectively, the edge states propagating along the nanoribbon edges are spin degenerate and valley polarized, while at the \textit{n-p} interface, valley degeneracy is preserved and four spin- and valley- degenerate channels co-propagate. In this regime, experimental works\cite{Williams2007,Lohmann2009,Matsuo2015bis,Klimov2015} show that the conductance is quantized to $e^2/h$, in agreement with theory\cite{Abanin2007} assuming complete mode mixing along the \textit{n-p} interface.\commentout{Calculations based on random matrix theory provides the same expectation value for the average conductance.\cite{XXX}}\\
\indent The microscopic mechanism at the origin of this mode mixing is however not clearly established. Numerical studies investigated the role of on-site disorder\cite{Li2008,Long2008} and edge/interface roughness.\cite{Low2009,Lagasse2016,Myoung2017} Semiclassical snakelike trajectories at the interface were considered as a possible source of mode mixing in the clean limit.\cite{Carmier2010,Carmier2011} The full quantum calculation reported in Ref.\cite{Tworzydlo2007} for the case of an ideal clean sample led to another prediction and pointed out the role of edge boundary conditions controlling the valley isospins of the valley polarized edge states. Experimental signatures of this effect were recently observed\cite{Handschin2017}.   
Finally, the crossover from the clean\cite{Tworzydlo2007} to the strongly disordered limit\cite{Abanin2007} has been investigated in Ref.\cite{Frassdorf2016}, still under the hypothesis of coherent transport. On the other hand, shot noise measurements were also reported in Refs.\cite{Matsuo2015,Kumada2015}. In particular, in Ref.\cite{Kumada2015}, the noise has been shown to vanish exponentially with the interface length. This behavior clearly suggests the existence of inelastic scattering along the interface, leading to energy relaxation and decoherence.

In this article, we investigate numerically the effect of incoherent scattering along the \textit{n-p} interface. We use fictitious voltage probes to model inelastic scattering, as proposed by B\"uttiker in Ref.\cite{Buttiker1986}. Though this model does not capture the microscopic origin of inelastic scattering, \textit{e.g.} electron-electron or electron-phonon interactions, it has been used in various contexts (see \textit{e.g.} Refs.\cite{Texier2000,Roulleau2009,Xing2008,Chen2011,Sanchez2016}) and has proven to be an efficient phenomenological technique for describing incoherent effects. In particular, the probe model was implemented in Ref.\cite{Chen2011} to study numerically the interplay between disorder and decoherence effects in the graphene \textit{n-p} junction at filling factors $(\nu_n,\nu_p)=(2,-2)$. Our study has some commonalities with Ref.\cite{Chen2011} but differs in two main points. First, we compute not only the conductance $G$ of the graphene \textit{n-p} junction, but also the noise $Q$ and its Fano factor $F$. In particular, we show that the conductance is not enough to probe the decoherence processes, making shot noise a fundamental quantity to unveil the \textit{n-p} junctions properties. 
Second, we use a scaling approach to extract the coherence length $L_\phi$ of the system. This allows us to encapsulate in a single parameter with clear physical meaning ($L_\phi$) the complex dependency of $G$, $Q$, and $F$ on various model parameters, notably on the virtual probe parameters. This approach makes the discussion of the experimental data much more straightforward. We study the behavior of $G$, $Q$, and $F$ with the interface length expressed in units of $L_\phi$, and eventually compare our numerical results to the experimental data reported in Ref.\cite{Kumada2015}.\\
\indent The paper is outlined as follows. In Sec.\,\ref{sec:model}, we introduce the scaled tight binding model of the graphene \textit{n-p} junction under perpendicular magnetic field, as well as the probe model. In Sec.\,\ref{sec_GandQcalc}, we explain how the conductance and the noise are calculated in the presence of the probes. The process of data analysis leading to the extraction of the coherence length is described in Sec.\,\ref{sec_scaling}. The results for the clean junction (without disorder) are given in Sec.\,\ref{sec_results_nodis}, from the coherent to the incoherent regimes. Disorder effects are discussed in Sec.\,\ref{sec_results_dis}. We conclude in Sec.\,\ref{ccl}.

\section{Model}
\label{sec:model}
\begin{figure}
    \includegraphics[clip,keepaspectratio,width=0.9\columnwidth]{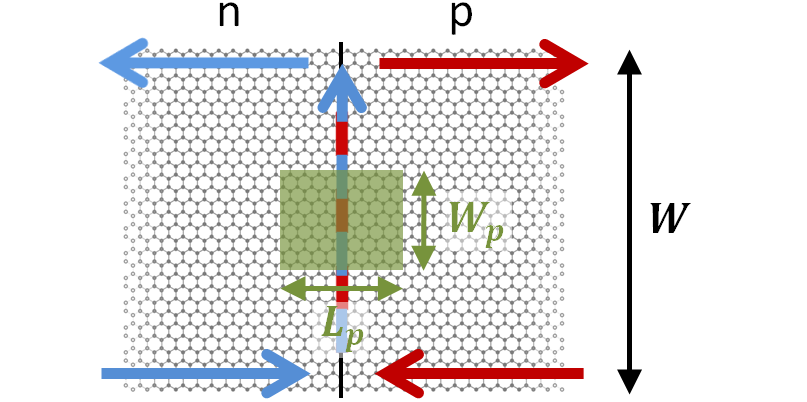}
    \caption{(Color online) Sketch of the graphene \textit{n-p} junction in the quantum Hall regime $(\nu_n,\nu_p)=(2,-2)$. The (blue) electron-like and (red) hole-like edge states propagate in opposite directions in the $n$ and $p$ part respectively. The incoming states at the bottom edge merge when they reach the \textit{n-p} junction, then mix along the interface (blue/red dashed line), and finally split when they reach the top edge. One-dimensional virtual probes are attached to the sites of the graphene layer located in the green rectangle of length $L_p$ and width $W_p$ in the middle of the \textit{n-p} interface. $W$ denotes the nanoribbon width.}
   \label{fig:sys}
\end{figure}

We consider a graphene nanoribbon of width $W$ connected to two left and right electronic reservoirs held at zero temperature. The system is described by the spinless tight binding Hamiltonian
\be
\label{eq_def_H}
H = -\sum_{\langle i,j \rangle}t e^{i\Phi_{ij}}c_i^\dagger c_j + \sum_i (V_i-\mu)c_i^\dagger c_i\,.
\ee
Twofold spin degeneracy will be restored in the conductance and noise formula in the next section. 
$c_i^\dagger$ and $c_i$ are respectively the electron creation and annihilation operators at site $\mathbf{r}_i=(x_i,y_i)$. The sum $\sum_{\langle i,j \rangle}$ is restricted to nearest neighbors. $t$ is the hopping term and $\Phi_{ij}=(e/\hbar)\int_{\mathbf{r}_j}^{\mathbf{r}_i}\mathbf{A}.d\mathbf{r}$ is the Peierls phase accounting for the presence of the perpendicular magnetic field $\mathbf{B}=\mathbf{\nabla} \times \mathbf{A}$. $\mu$ is a constant potential applied everywhere and $V_i=V(\mathbf{r}_i)$ is a (gate-defined) step potential,
\be
\label{eq_dfsteppot}
V(\mathbf{r}_i) = \frac{V_g}{2}\left[ 1+\tanh\left( \frac{2x_i}{l}\right)\right]\,,
\ee
interpolating from $0$ in the left part ($x<0$) to $V_g$ in the right part ($x>0$), over a characteristic length $l$.\\
\indent To save computation time, we consider the scaled model of graphene with nearest-neighbor hopping term $t=t_0/s_f$ and lattice spacing $a=s_fa_0$, $a_0=0.142$\,nm and $t_0=2.8$\,eV being the lattice spacing and hopping term for real graphene, and $s_f$ a scaling parameter. It has been shown in Ref.\cite{Liu2015} that low energy transport properties of real graphene can be captured by this scaled model as long as $s_f\ll 3t_0\pi/|E_{\mathrm{max}}|, l_B/a_0$, $E_{\mathrm{max}}$ being the maximal energy of interest and $l_B=\sqrt{\hbar/(eB)}$ the magnetic length.\\
\indent To mimic decoherence, we add fictitious voltage probes as follows. In a rectangular region of width $W_p$ and length $L_p$ centered around $x=0$ (see Fig.\ref{fig:sys}), each site of the graphene layer is attached to a semi-infinite one-dimensional chain -- a probe -- with zero on-site energy and nearest-neighbor hopping term $t$. We note\footnote{We used this definition but did not investigate scaling properties with $s_f$.} $t_p/s_f$ the hopping term between the site in the graphene layer and the neighbor site in the chain. At the reference energy $E=0$ at which transport is investigated in the following, the self-energy of a probe is purely imaginary and reads $\Sigma_p=-it_p^2/t/s_f^2$. Note that for computational reasons, we also consider the situation where only a finite fraction $\alpha_p$ of carbon atoms in the $W_p \times L_p$ rectangle, chosen randomly, are effectively attached to probes. We will see in Sec.\,\ref{sec_scaling} that the choice of $\alpha_p$ is physically irrelevant.\\
\indent Throughout the paper, we take $s_f=10$, $B=10$\,T, $\mu=0.05$\,eV and $V_g=0.11$\,eV. Since the energy gap between the zero-th and first Landau level is of the order of $0.1$\,eV for this value of $B$, the filling factors in the \textit{n} and \textit{p} regions are $\nu_n=2$ and $\nu_p=-2$ respectively. The length scale of the \textit{n-p} junction\footnote{It is controlled by the thickness of the insulating layers between the graphene sheet and the top gate in Ref.\cite{Kumada2015}.)} is fixed to $l=5$\,nm $\gtrsim a$ and the length $L_p$ of the region where probes are attached is fixed to $L_p=30$\,nm $\gtrsim l_B\approx 8.2$\,nm to cover (along the x-direction) the spatially superimposed interface states centered around $x=0$. In most cases, the width of the graphene ribbon is fixed to $W\approx 160$\,nm while $W_p$ is varied. In the following, we denote the number of hexagons across the ribbon by $\mathcal{N}$. Under the zigzag [armchair] boundary condition, $W$ is related to $\mathcal{N}$ by $W=(3\mathcal{N}+1)a/2$  $[W=\mathcal{N}\sqrt{3}a]$.

\section{Conductance and noise calculation in the presence of the probes}
\label{sec_GandQcalc}
The total system made of the graphene layer connected to the left and right leads $L$ and $R$ and the probes $p=1,...,N_p$ is supposed to be phase coherent.\commentout{We start by solving numerically the coherent scattering problem with the KWANT software~\cite{Groth2014}.} We describe its transport properties within the standard Landauer-B\"uttiker formalism.\cite{Datta1995} To introduce inelastic scattering, we follow the approach introduced by B\"uttiker in Ref.\cite{Buttiker1986} and reviewed in Ref.\cite{Blanter2000}. We impose the current $I_p$ flowing through each fictitious probe $p$ to be zero: an electron that comes out into the lead $p$ is eventually absorbed in the reservoir and has to be replaced by another electron injected from the lead $p$. Since the phases and the energies of the two electrons are uncorrelated, inelastic decoherence effects are induced in the system. Thereby, the coherent problem with $N_p+2$ reservoirs reduces to an effective incoherent problem with only two left and right reservoirs. 

\commentout{We use the KWANT software~\cite{XXX} to compute the set of transmission coefficients $T_{\alpha\beta}$ from the lead $\beta$ to the lead $\alpha$ at the Fermi energy $E_F=0$ ($\alpha=L,R$ or $1,...,N_p$). 
We work at zero temperature and investigate transport through the graphene layer around the Fermi energy $E_F=0$. Without loss of generality, we assume that a small voltage bias $V/2$ is applied on the left lead $L$ and $-V/2$ on the right lead $R$. This generates electric currents $I_\alpha$ flowing from the leads $\alpha$ ($\alpha=L,R$ or $p=1,...,N_p$). Following the approach introduced by B\"uttiker in Ref.\cite{Buttiker1986}, we impose that the average current $I_p$ flowing through each fictitious probe $p=1,...,N_p$ is zero. From the condition $I_p=0$, we find the voltages $V_p$ at all probes $p$. In practice, this requires to solve the linear system}

We work at zero temperature and investigate transport through the graphene layer around the Fermi energy $E_F=0$. Without loss of generality, we assume that a small voltage bias $V/2$ is applied on the left lead $L$ and $-V/2$ on the right lead $R$. This generates electric currents $I_\alpha$ flowing from the leads $\alpha$ ($\alpha=L,R$ or $p=1,...,N_p$). To compute the conductance of the effective two-terminal problem, it is enough to impose a zero average current $\left\langle I_p \right\rangle=0$ in all probes $p$. Therefore, in virtue of current conservation, $\left\langle I_L \right\rangle = -\left\langle I_R \right\rangle$. From the condition $\left\langle I_p \right\rangle=0$, we find the voltages $V_p$. In practice, this requires to solve the linear system
\begin{equation}
\label{eq_linearsystemVq}
\left\langle I_p \right\rangle=-\sum_{q\,\in\,\mathcal{P}}A_{pq}V_q+B_p=0~~~\forall\,p\in\mathcal{P}
\end{equation}
where $\mathcal{P}=\lbrace1,...,N_p\rbrace$, $A_{pp}=T_{pp}-1$, $A_{pq}=T_{pq}$ if $p\neq q$ and $B_p=(T_{pR}-T_{pL})V/2$. Here $T_{\alpha\beta}$ denotes the transmission probabilities from the lead $\beta$ to the lead $\alpha$ (or reflection probabilities if $\alpha=\beta$). They are calculated with the KWANT software (see Ref.\cite{Groth2014} and footnote\footnote{See \url{https://kwant-project.org/}.}). The conductance $G=\left\langle I_L \right\rangle / V$ follows immediately 
\begin{equation}
\label{eq_GwithProbes}
G=\frac{2e^2}{h}\left[ T_{LR} + \sum_{p\,\in\,\mathcal{P}}\, T_{Lp}\left(\frac{1}{2}-\frac{V_p}{V}\right)\right]\,.
\end{equation}
The factor $2$ in Eq.\eqref{eq_GwithProbes} accounts for the spin degeneracy.

Let us proceed with the noise calculation. We now impose\cite{Beenakker1992,Blanter2000} that the currents $I_p(t)$ in the probes vanish at each instant of time $t$. The voltages $V_p(t)$ at the probes become fluctuating and are assumed to adjust instantaneously to ensure that $I_p(t)=0$. This is justified as long as transport properties are investigated at low frequency~\cite{Beenakker1992}. To compute the noise, we first write the currents as
\begin{align}
I_L(t) & = T_{LR}V + \sum_{p\,\in\,\mathcal{P}}T_{Lp}\left(\frac{V}{2}-V_p(t)\right)+\delta I_L(t) \label{eq_ILoft}\\
I_R(t) & = -T_{RL}V - \sum_{p\,\in\,\mathcal{P}}T_{Rp}\left(\frac{V}{2}+V_p(t)\right)+\delta I_R(t) \\
I_p(t) & =-\sum_{q\,\in\,\mathcal{P}}A_{pq} V_q(t)+B_p+\delta I_p(t)~~~\forall\,p\in\mathcal{P} \label{eq_Ipoft}
\end{align}
and introduce thereby the current fluctuations $\delta I_L$, $\delta I_R$ and $\delta I_p$ ($p\in\mathcal{P}$). Each of them is zero on average and the two point correlators 
\begin{equation}
\label{eq_dfrawnoise}
P_{\alpha\beta}\equiv 2\int\mathrm{d}t\left\langle \delta I_\alpha(t) \delta I_\beta(0) \right\rangle
\end{equation}
are given at zero temperature by the formula~\cite{Buttiker1992,Blanter2000}
\begin{equation}
\label{eq_rawnoise}
\begin{split}
P_{\alpha\beta}= \frac{2e^2}{h} \sum_{i\neq j}\int\mathrm{d}E & \Bigl\{\mathrm{Tr}\left[S_{\alpha i}^\dagger S_{\alpha j} S_{\beta j}^\dagger S_{\beta i} \right]\\
&\times\left[f_i(1-f_j)+f_j(1-f_i)\right]\Bigr\}\,.
\end{split}
\end{equation}
Note that Eqs.\,\eqref{eq_dfrawnoise} and \eqref{eq_rawnoise} are valid for all leads $\alpha$ and $\beta$ (the probes 	and the left and right leads). Also, the sum in Eq.\,\eqref{eq_rawnoise} runs over all leads. $f_i(E)=\theta(\mu_i-E)$ is the zero temperature Fermi distribution of the reservoir $i$ with $\mu_L=eV/2$, $\mu_R=-eV/2$ and $\mu_p=e\left\langle V_p \right\rangle$ for the probes $p$. $S_{\alpha j}$ denotes the scattering matrix element from the lead $j$ to the lead $\alpha$. We use the KWANT software~\cite{Groth2014} to compute the scattering matrix $S$ at the Fermi energy $E_F=0$ and perform the energy integral in Eq.\eqref{eq_rawnoise} upon neglecting the energy dependency of $S$ around $E_F$ ($S(E)\approx S(0)$). Thus we compute the correlators $P_{\alpha\beta}$.
 
We now introduce the two-terminal zero-frequency noise $Q_{\alpha\beta}$ defined for $\alpha,\beta=L$ or $R$ as
\begin{equation}
\label{eq_dfnoise}
Q_{\alpha\beta}\equiv 2\int\mathrm{d}t\left\langle \Delta I_\alpha(t) \Delta I_\beta(0) \right\rangle
\end{equation}
where $\Delta I_\alpha(t)\equiv I_\alpha(t)-\left\langle I_\alpha \right\rangle$ denotes the current fluctuations. \commentout{In the low frequency limit, current fluctuations are conserved \textit{i.e.} the sum of current fluctuations over all leads vanish.}Since $I_p(t)=0$ in probes $p$, $\Delta I_L(t)+\Delta I_R(t)=0$. Therefore, $Q_{LL}=Q_{RR}=-Q_{LR}=-Q_{RL}$. To compute \textit{e.g.} $Q_{LL}$, we write (using Eq.\eqref{eq_ILoft})
\begin{equation}
\Delta I_L(t) = -\sum_{p\,\in\,\mathcal{P}} T_{Lp}\,\Delta V_p(t) + \delta I_L(t)
\end{equation}
and express the voltage fluctuations $\Delta V_p(t)\equiv V_p(t)-\left\langle V_p \right\rangle$ as 
\begin{equation}
\label{eq_Vpoft}
\Delta V_p(t)=\sum_{q\,\in\,\mathcal{P}}A^{-1}_{pq}\,\delta I_q(t)~~~\forall\,p\in\mathcal{P}
\end{equation}
by imposing the condition $I_p(t)=0$ in Eq.\eqref{eq_Ipoft}. Finally, we deduce from Eqs.\eqref{eq_dfrawnoise} and \eqref{eq_dfnoise}-\eqref{eq_Vpoft}
\begin{equation}
\begin{split}
Q_{LL}=\sum_{p,\tilde{p},q,\tilde{q}\,\in\,\mathcal{P}}T_{Lp}\,T_{L\tilde{p}}\,A^{-1}_{pq}\,A^{-1}_{\tilde{p}\tilde{q}}\,P_{q\tilde{q}}\\
-2 \sum_{p,q\,\in\,\mathcal{P}}T_{Lp}\,A^{-1}_{pq}\,P_{qL}+P_{LL}\,.
\end{split}
\end{equation}
Hereafter, we note $Q\equiv Q_{LL}$ and express the noise $Q$ in units of $Q_0\equiv 2e^3V/h$. The conductance $G$ is given in units of $G_0\equiv 2e^2/h$. We also compute the dimensionless Fano factor $F\equiv Q/(2eGV)=(Q/Q_0)/(2G/G_0)$.

\section{Scaling approach}
\label{sec_scaling}
\begin{figure}
    \includegraphics[clip,keepaspectratio,width=\columnwidth]{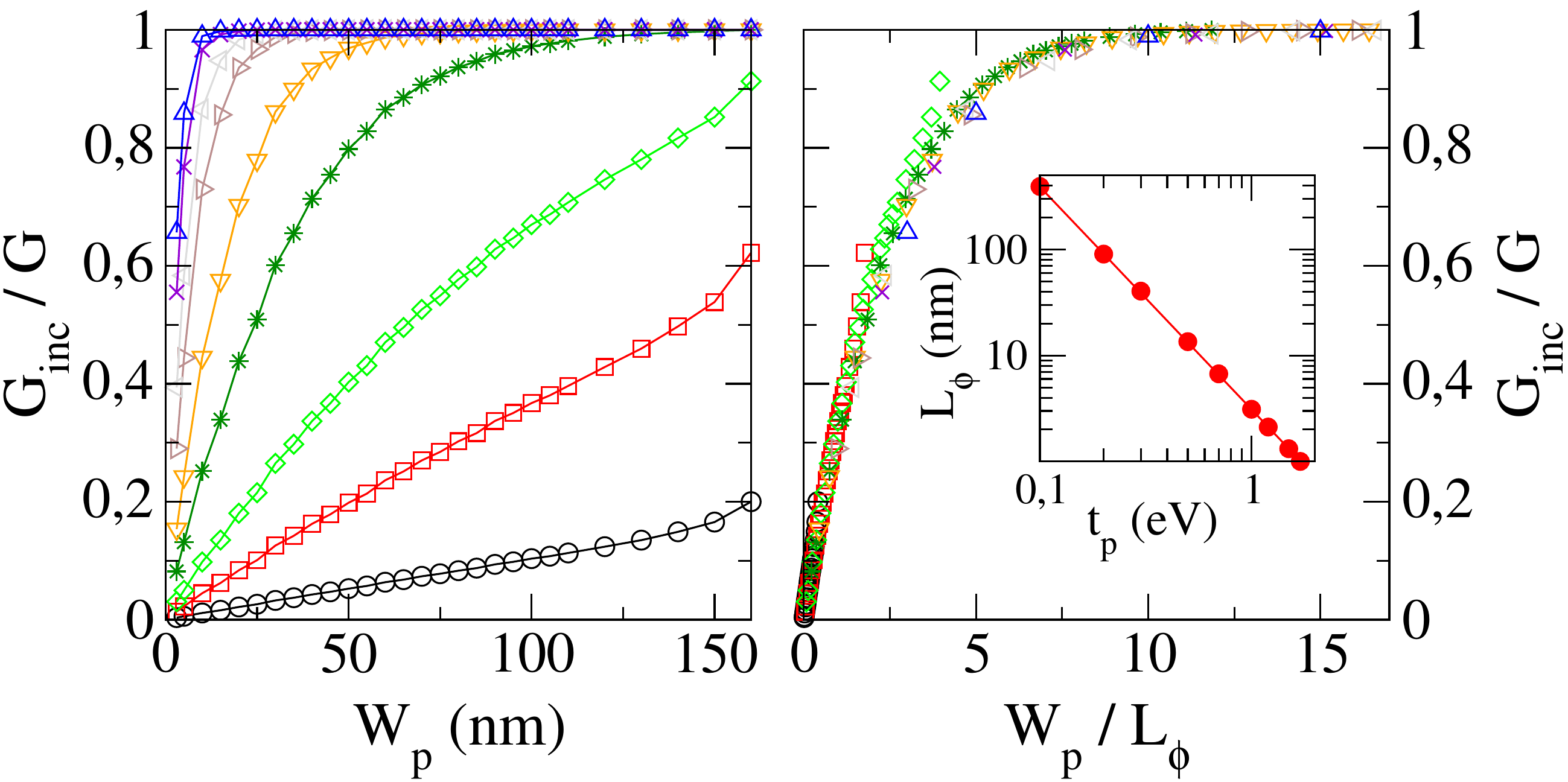}
    \caption{(Color online) (Left panel) $G_{\mathrm{inc}}/G$ as a function of the width $W_p$ for different values of the system-probe coupling $t_p$ ($t_p=0.1$ ({\color{black}{\scriptsize{\Circle}}}), $0.2$ ({\scriptsize{\color{red}$\square$}}), $0.3$ ({\small{\color{green}$\Diamond$}}), $0.5$ ({\normalsize{\color{darkgreen}$\ast$}}), $0.7$ ({\small{\color{orange}$\triangledown$}}), $1$ ({\small{\color{brown}$\triangleright$}}), $1.2$ ({\small{\color{gray}$\triangleleft$}}), $1.5$ ({\small{\color{violet}$\times$}}), and $1.7$\,eV ({\scriptsize{\color{blue}$\triangle$}})). Lines are guides to the eye. We took $\alpha_p=1$ and zigzag edges with $\mathcal{N}=74$ (\textit{i.e.} $W=158.33$\,nm). (Right panel) Same data after rescaling along the $x$-axis. The dependence on $t_p$ of the scaling parameter $L_{\phi}(t_p)$ is given in the inset. The red line is a fit $L_{\phi}=3.1/t_p^{2.1}$.}
   \label{fig:Ginc_scaling}
\end{figure}

To investigate decoherence effects on the conductance $G$ and the noise $Q$ of the graphene \textit{n-p} junction, we need to compute the two quantities for different values of the width $W_p$ and of the hopping term $t_p$ (introduced in Sec.\,\ref{sec:model}). To analyze our data, we use a scaling procedure illustrated in Fig.\ref{fig:Ginc_scaling}. We consider the incoherent contribution $G_\mathrm{inc}$ to the total conductance $G$. It corresponds to electrons that flow indirectly from the left lead to the right one via the probes. $G_\mathrm{inc}$ is given by the second term in the right hand side of Eq.\,\eqref{eq_GwithProbes}. As shown in the left panel of Fig.\ref{fig:Ginc_scaling}, the ratio $G_\mathrm{inc}/G$ increases when the coupling $t_p$ to the probes is increased or when the region covered by the probes is made larger (by increasing $W_p$). In the right panel of Fig.\ref{fig:Ginc_scaling}, we show that the curves of $G_\mathrm{inc}/G$ versus $W_p$ for different $t_p$ can all be superimposed on top of each other if for each $t_p$, one rescales by hand the $x$-axis $W_p$ to $W_p/L_\phi(t_p)$. The extracted scaling parameter $L_\phi(t_p)$ can be interpreted as the coherence length of the system.\footnote{Following Ref.\cite{Xing2008}, we also defined for each $t_p$ another coherence length $\tilde{L}_\phi(t_p)$ as the value of $W_p$ for which $G_\mathrm{inc}/G=1/2$. When plotted as a function of $W_p/\tilde{L}_\phi(t_p)$, the data of Fig.\ref{fig:Ginc_scaling} also collapse onto one single curve and we find $L_\phi\approx 0.48\tilde{L}_\phi$. Hence the two procedures used to extract the system coherence length are consistent with each other.} It is defined up to a multiplicative constant depending on the curve we choose as the reference for rescaling the other curves. We have taken $L_\phi=1$\,nm for $t_p=1.7$\,eV in Fig.\ref{fig:Ginc_scaling}.

A closer look at the right panel of Fig.\ref{fig:Ginc_scaling} reveals that the scaling actually breaks down when $W_p$ approaches $W$ (see \textit{e.g.} the green diamonds). This is due to the fact that when $W_p\approx W$, the probes are attached up to the extremities of the \textit{n}-\textit{p} interface and therefore modify scattering processes at the top and bottom corners of width $\sim l_B$ where edge and interface modes meet. We have checked that the discrepancy fades out for large $W\approx W_p$ when the contribution of this corner effect becomes negligible.

\begin{figure}
    \includegraphics[clip,keepaspectratio,width=\columnwidth]{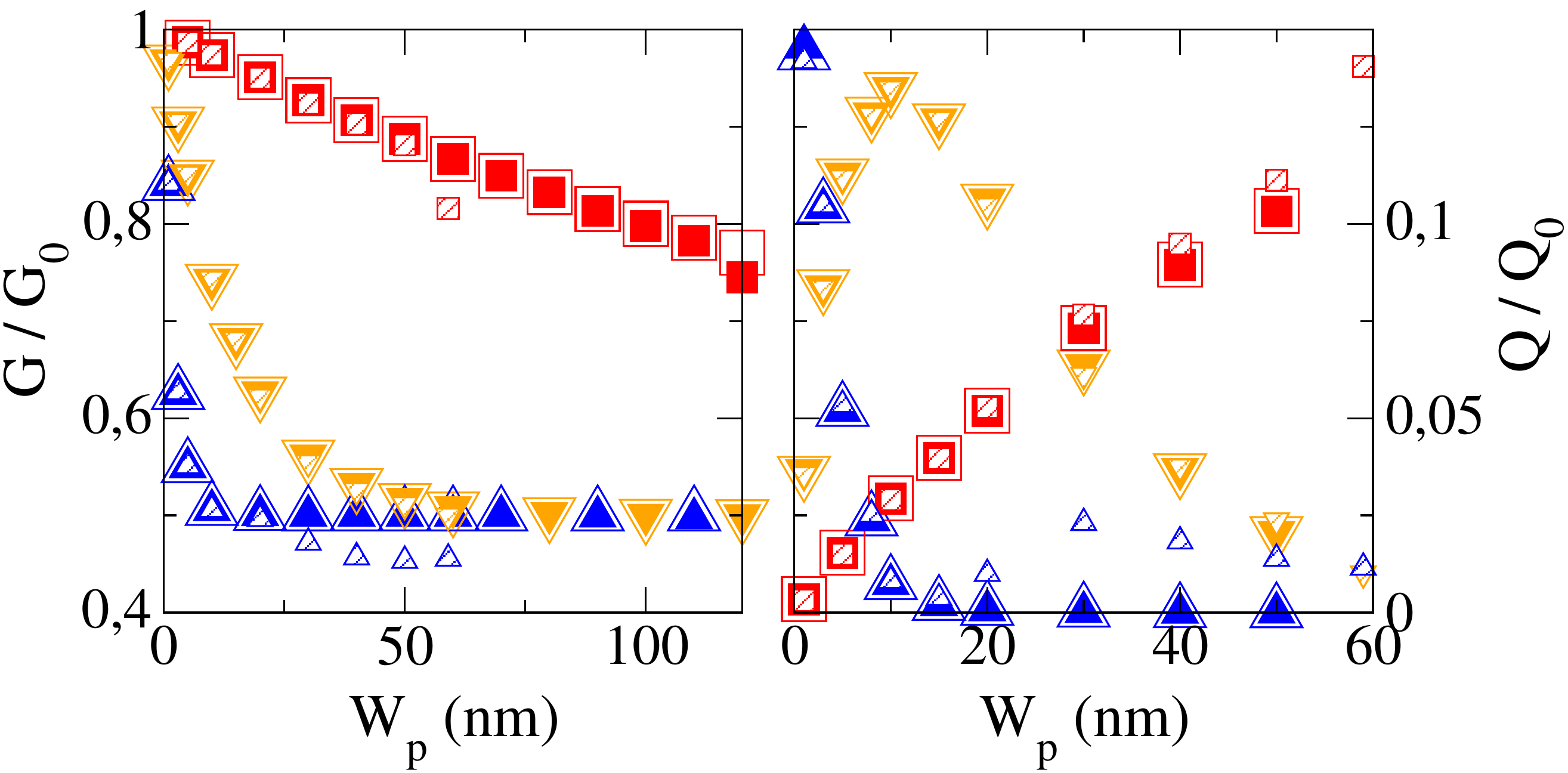}   
    \caption{(Color online) Conductance $G/G_0$ (left) and noise $Q/Q_0$ (right) as a function of $W_p$ for different widths $W=(3\mathcal{N}+1)a/2$ of a zigzag nanoribbon ($\mathcal{N}=26$ (striped symbols), $56$ (full symbols), and $74$ (empty symbols)). Data are plotted with $\alpha_p=1$ for $t_p=0.2$ ({\scriptsize{\color{red}$\square$}}), $0.7$ ({\small{\color{orange}$\triangledown$}}), and $1.7$\,eV ({\scriptsize{\color{blue}$\triangle$}}). Finite size effects in $W$ are visible when $W_p\approx W$.}
   \label{fig:G_Noise_diffW}
\end{figure}

\begin{figure}
    \includegraphics[clip,keepaspectratio,width=\columnwidth]{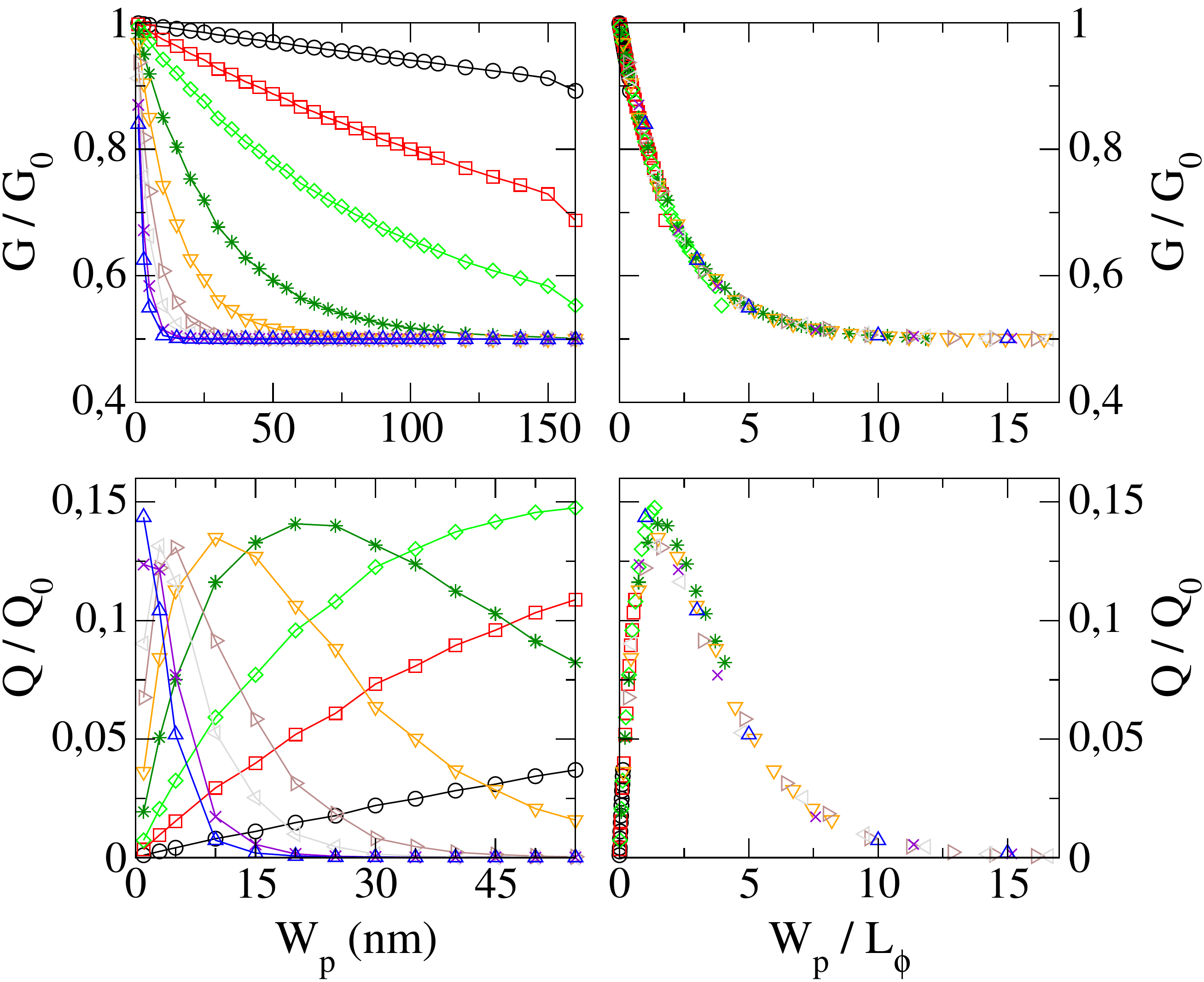}
    \caption{(Color online) (Left panel) Conductance $G/G_0$ (top) and noise $Q/Q_0$ (bottom) as a function of the width $W_p$ for different values of the system-probe coupling $t_p$ ($t_p=0.1$ ({\color{black}{\scriptsize{\Circle}}}), $0.2$ ({\scriptsize{\color{red}$\square$}}), $0.3$ ({\small{\color{green}$\Diamond$}}), $0.5$ ({\normalsize{\color{darkgreen}$\ast$}}), $0.7$ ({\small{\color{orange}$\triangledown$}}), $1$ ({\small{\color{brown}$\triangleright$}}), $1.2$ ({\small{\color{gray}$\triangleleft$}}), $1.5$ ({\small{\color{violet}$\times$}}), and $1.7$\,eV ({\scriptsize{\color{blue}$\triangle$}})). Lines are guides to the eye. We took $\alpha_p=1$ and zigzag edges with $\mathcal{N}=74$ (\textit{i.e.} $W=158.33$\,nm). (Right panel) Same data plotted as a function of $W_p/L_\phi$ using the scaling parameter $L_{\phi}(t_p)$ extracted in Fig.\ref{fig:Ginc_scaling}. All $G$ data turn out to collapse on one single curve. This also holds for $Q$.}
   \label{fig:G_Noise_scaling}
\end{figure}

We now turn to the study of the conductance $G$ and the noise $Q$. In continuity with the previous remark, we first note from Fig.\ref{fig:G_Noise_diffW} that $G$ and $Q$ are independent of $W\geq W_p$ or in other words do not depend on the length $W-W_p$ of the interface region not covered by probes. Finite size effects (originating from the corner effect discussed just before) are nevertheless visible at small $W$ when $W_p\approx W$. This is the reason why hereafter, we investigate the dependency on the interface length by varying $W_p$ at fixed $W\gg W_p$. Then, we show in Fig.\ref{fig:G_Noise_scaling} that the coherence lengths $L_\phi(t_p)$ extracted previously by rescaling the curves of $G_\mathrm{inc}/G$ can also be used to rescale in the same way the curves of $G$ and $Q$. Actually, this scaling procedure also works if the magnetic field $B$ is varied and $L_\phi$ is made $B$-dependent. This also holds for the probe filling rate $\alpha_p$. Thus, we find in the end that 
\begin{equation}
\label{eq_scaling}
A\left(W,W_p,t_p,\alpha_p,B \right) = A\left(\frac{W_p}{L_{\phi}(t_p,\alpha_p,B)}\right)
\end{equation}
for $A=G_\mathrm{inc}/G$, $G$, $Q$ or $F$. This is true up to the (small) finite size effects mentioned above. Eq.\,\eqref{eq_scaling} tells us that the entire curves $A(W_p/L_\phi)$ can be determined by varying either $W_p$ or one of the parameters $t_p$, $\alpha_p$ or $B$. To be more precise, $B$ can only be varied in a small range of values to preserve the condition $s_f\ll l_B/a_0$ (see Sec.\ref{sec:model}) and to remain in the regime where $\nu_n=2$ and $\nu_p=-2$. Hence, varying $B$ only give us access to a small part of the curves $A(W_p/L_\phi)$ if all other parameters are fixed. Besides, taking $\alpha_p<1$ would require in principle to average over the different spatial distribution of the probes in the probe region. To avoid this time-consuming step, we only consider values $0.25\leq \alpha_p\leq 1$ for which the variations of $A$ from one probe configuration to another is negligible. Apart from these technical considerations, our scaling approach summarized by Eq.\,\eqref{eq_scaling} results in the elimination of the parameters $t_p$ and $\alpha_p$ of the fictitious probe model, by encapsulating them in the coherence length $L_\phi$. This allows us to bridge the gap between our model and the realistic problem and eventually to study how the quantities $G$, $Q$ and $F$ behave when the length of the graphene \textit{n-p} interface is varied with respect to the system coherence length.  

\section{Results without disorder}
\label{sec_results_nodis}
\begin{figure}
    \includegraphics[clip,keepaspectratio,width=0.95\columnwidth]{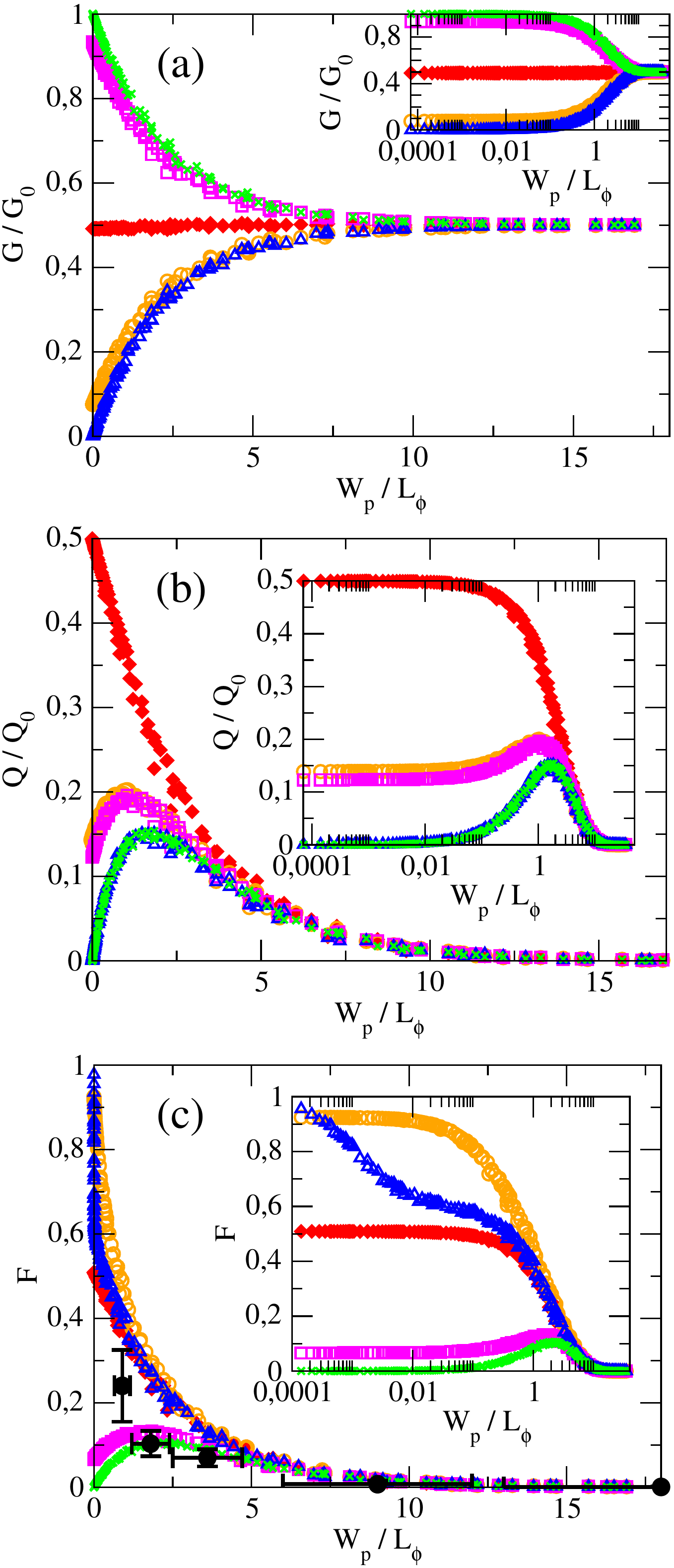}
    \caption{(Color online) Main panels: Conductance $G/G_0$ (a), noise $Q/Q_0$ (b) and Fano factor $F$ (c) of the clean junction as a function of $W_p/L_\phi$, for different kinds of edges (zigzag A, $\mathcal{N}=74$ ({\scriptsize{\color{green}$\times$}}); zigzag B, $\mathcal{N}=75$ ({\scriptsize{\color{blue}$\triangle$}}); armchair A, $\mathcal{N}=65$ ({\scriptsize{\color{magenta}$\square$}}); armchair B, $\mathcal{N}=64$ ({\color{orange}{\scriptsize{\Circle}}}); and armchair C, $\mathcal{N}=66$ ({\scriptsize{\color{red}\ding{117}}})). Each curve has been obtained with $\alpha_p=0.25$, by varying $W_p$ from $10$ to $150$\,nm and $t_p$ from $0.01$ to $1.7$\,eV, upon rescaling data using $L_\phi(t_p)$ as explained in Sec.\,\ref{sec_scaling}. The black dots in (c) correspond to experimental data issued from Ref.\cite{Kumada2015}. Vertical error bars are the experimental ones. Horizontal error bars are due to the mapping from the experimental to the numerical units along the $x$-axis (see text). Insets: same data with $x$-axis in log-scale.}
   \label{fig:alldata_vs_Wp}
\end{figure}

The conductance $G$ of the graphene \textit{n-p} junction in the regime $(\nu_n,\nu_p)=(2,-2)$ has been calculated in Ref.\cite{Tworzydlo2007} in the clean and coherent limit.\commentout{In the absence of intervalley scattering, it is determined by the valley isospins of the two (spin-degenerate) valley-polarized edge states at the opposite edges of the graphene ribbon.\cite{Tworzydlo2007} Therefore, it depends on the boundary conditions.} In the case of ribbons with armchair edges, $G$ turns out to depend on the number $\mathcal{N}$ of hexagons across the ribbon modulo three and on the parameters $V_g$, $\mu$, and $B$ (defined in Sec.\ref{sec:model}). In the limit of large $V_g-\mu$ (but still in the regime $(\nu_n,\nu_p)=(2,-2)$), $G/G_0 \to 1$ if $\mathcal{N}\,\mathrm{mod}\,3 = 2$ and $G/G_0 \to 0.25$ otherwise.\footnote{Note that the convention used for building the armchair ribbons are not the same in Ref.\cite{Tworzydlo2007} and in the present paper. In Ref.\cite{Tworzydlo2007} (though not explicitly mentioned), the number $\mathcal{N}$ of hexagons across the ribbon is fixed whatever the position along the ribbon direction. In the present paper, the number of hexagons switches from  $\mathcal{N}$ to $\mathcal{N}-1$ to $\mathcal{N}$ to $\mathcal{N}-1$ etc ... along the ribbon direction. In other words, $\mathcal{N}$ is related to the ribbon width $W$ by $W=\mathcal{N}\sqrt{3}a$ in our case and by $W=(\mathcal{N}+1/2)\sqrt{3}a$ in Ref.\cite{Tworzydlo2007}. This explains why for instance $G\approx G_0$ for  $\mathcal{N}\,\mathrm{mod}\,3 = 1$ in Fig.5 of Ref.\cite{Tworzydlo2007} and for $\mathcal{N}\,\mathrm{mod}\,3 = 2$ in Fig.\ref{fig:alldata_vs_Wp}(a) of the present paper (see square symbols for $W_p/L_\phi \ll 1$).} For intermediate $V_g-\mu$, $G/G_0$ values for $\mathcal{N}\,\mathrm{mod}\,3 = 0$ and $1$ are different but average out at $0.25$ approximately. In the case of zigzag edges, it is found that $G=0$ if $\mathcal{N}$ is odd while $G=G_0$ if $\mathcal{N}$ is even, independently of other parameters as long as $(\nu_n,\nu_p)=(2,-2)$.
In the following, we will consider five kinds of edges: \textit{(i)} zigzag edges with even $\mathcal{N}$ (labeled zigzag A), \textit{(ii)} zigzag edges with odd $\mathcal{N}$ (labeled zigzag B), \textit{(iii)} armchair edges with $\mathcal{N}\,\mathrm{mod}\,3 = 2$ (labeled armchair A), \textit{(iv)} armchair edges with $\mathcal{N}\,\mathrm{mod}\,3 = 1$ (labeled armchair B), and \textit{(v)} armchair edges with $\mathcal{N}\,\mathrm{mod}\,3 = 0$ (labeled armchair C). 

\commentout{Let us now study decoherence effects on the transport properties of the clean \textit{n-p} junction.}In Fig.\ref{fig:alldata_vs_Wp}(a), we have plotted for each of the five types of edges \commentout{introduced above}the conductance $G$ of the clean \textit{n-p} junction as a function of $W_p/L_\phi$ following the approach described in Sec.\,\ref{sec_GandQcalc} and Sec.\,\ref{sec_scaling}. Let us first discuss our results in the coherent limit ($W_p/L_\phi \ll 1$). When zigzag edges are considered, we find that $G=G_0$ for even $\mathcal{N}$ (zigzag A) and $G=0$ for odd $\mathcal{N}$ (zigzag B), in agreement with Refs.\cite{Tworzydlo2007,Akhmerov2008}. In the case of armchair edges, $G$ is determined by our choice of $V_g$, $\mu$, and $B$ parameters (fixed at the end of Sec.\ref{sec:model}). We find $G/G_0\approx 0.93$ for armchair A edges, $G/G_0\approx0.075$ for armchair B edges and $G/G_0\approx0.49$ for armchair C edges: Qualitatively, $G/G_0$ is close to $1$ for armchair A edges and the average value of the other two values of $G/G_0$ for armchair B and C edges is close to $0.25$.\footnote{Note that the theoretical limit of large $V_g-\mu$ where $G/G_0 \to 1$ for armchair A edges and $G/G_0 \to 0.25$ for armchair B and C edges cannot be reached here, with $B=10$\,T, without leaving the bipolar regime $(\nu_n,\nu_p)=(2,-2)$ corresponding to $0\leq V_g-\mu \lesssim 0.1$\,eV.} This qualitative statement remains true for another choice of $V_g$, $\mu$, and $B$ parameters even though the values of the conductance for the three armchair cases are modified. This is consistent with Fig.5 of Ref.\cite{Tworzydlo2007}.

When $W_p/L_\phi$ is increased, the edge-dependent features diminish and eventually in the incoherent limit ($W_p/L_\phi \gg 1$), we recover $G=G_0/2$ for all kinds of edges. This corresponds to the conductance plateau measured experimentally\cite{Williams2007,Lohmann2009,Matsuo2015bis,Klimov2015} when $(\nu_n,\nu_p)=(2,-2)$. This value also coincides with the original theoretical prediction of Abanin and Levitov~\cite{Abanin2007}, rederived in Refs.\cite{Li2008,Frassdorf2016} and confirmed by various numerical works~\cite{Li2008,Long2008,Low2009,Chen2011,Lagasse2016,Myoung2017} using different disorder models.\footnote{Note that contrary to Ref.\cite{Chen2011} where a probe model was also used, we obtain $G=G_0/2$ in the incoherent limit even without disorder. We believe the discrepancy is due to the fact that we attach probes to all sites of the graphene layer located in the rectangle of width $W_p$ and length $L_p$ at the $n$-$p$ interface (see Fig.\ref{fig:sys}) while in Ref.\cite{Chen2011}, probes are only attached along vertical lines $x=\pm\, l/2$ ($x=0$ corresponding to the $n$-$p$ interface and $l$ being the characteristic length of the potential step, $l>l_B$ in Ref.\cite{Chen2011}). The interface modes are therefore less sensitive to the presence of the probes and disorder is needed to enhance the effect of the probes.} However, the fact that various models lead to the same prediction does not allow us to identify the relevant physical mechanisms in play in experiments. This is the reason why we also study hereafter the noise $Q$ of the $n$-$p$ junction.\\
\indent The plots yielding $Q$ as a function of $W_p/L_\phi$ are shown in Fig.\ref{fig:alldata_vs_Wp}(b). In the coherent limit ($W_p/L_\phi \ll 1$), our data are in perfect agreement with the zero temperature shot noise formula\cite{Buttiker1992,Jong1997}
\begin{equation}
\label{eq_noise_noprobe}
\frac{Q}{Q_0}=2\,\frac{G}{G_0}\left(1-\frac{G}{G_0}\right)
\end{equation}
using the values of $G/G_0$ computed in Fig.\ref{fig:alldata_vs_Wp}(a) for $W_p/L_\phi \ll 1$.
In the opposite limit ($W_p/L_\phi \gg 1$), the noise $Q$ is found to vanish in all cases. This feature is consistent with the experimental observation of a suppressed shot noise for long interface lengths\,\cite{Kumada2015} and cannot be reproduced with a coherent disordered model.\cite{Frassdorf2016} It is a signature of incoherent mixing between interface modes which proves to play a crucial role in experiments.\cite{Kumada2015} Further analysis of the curves $Q(W_p/L_\phi)$ at large $W_p/L_\phi$ up to $35$ shows us that the noise $Q$ decreases exponentially with $W_p/L_\phi$.
Besides, we note that the curves $Q(W_p/L_\phi)$ are identical whether zigzag A or zigzag B edges are considered. The (quasi) superposition of the curves for armchair A and armchair B edges is however a coincidence due to the choice of $V_g$, $\mu$, and $B$ parameters.\\ 
\indent Finally, we show in Fig.\ref{fig:alldata_vs_Wp}(c) our numerical results for the Fano factor $F$, together with the experimental data reported in Fig.3b of Ref.\cite{Kumada2015}. In this paper, $F$ was measured at a temperature $4.2$\,K for different lengths $W$ of the \textit{n-p} interface ranging from $5$ to $100$ microns. For each length $W$, measurements were repeated for different values of the magnetic field (around $B=10$\,T) and for different values of $V_g$ so as to remain in the regime $(\nu_n,\nu_p)=(2,-2)$. Then, for each $W$, the mean value $\left\langle F\right\rangle$ of the Fano factor and its standard deviation $\sigma_F=(\left\langle F^2\right\rangle-\left\langle F\right\rangle^2)^{1/2}$ were extracted (by averaging over different values of $B$ and $V_g$). They are shown by black dots and vertical error bars in Fig.\ref{fig:alldata_vs_Wp}(c). The coherence length $l_\phi\approx 15\,\mu$m was finally estimated with an exponential fit $\left\langle F\right\rangle(W)\sim \exp(-W/l_\phi)$. The comparison between numerical data and experimental ones is hindered by the fact that the numerical coherence length $L_\phi$ is defined up to a multiplicative constant (see Sec.\,\ref{sec_scaling}). We proceed as follows. We fit the numerical Fano factor as $F\sim \exp(-W_p/(cL_\phi))$ at large $W_p/L_\phi$ and find $c\approx 2.7\pm 0.3$. Identifying\footnote{In the numerics, we vary $W_p$ at fixed $W\geq W_p+2\,l_B$ to avoid finite size effects mentioned in Sec.\,\ref{sec_scaling}.} $W$ with $W_p$, we get $l_\phi=c L_\phi$. The experimental values $\left\langle F\right\rangle(W)$ with their vertical error bars (standard deviations) are finally plotted in Fig.\ref{fig:alldata_vs_Wp}(c) as a function of $Wc/l_\phi$ using $l_\phi\approx 15\pm 3\,\mu$m. The horizontal error bars account for the rough estimations of $l_\phi$ and $c$. This procedure -- though not very accurate -- avoids using any adjustable parameter. However, it has to be stressed that the numerical data shown in Fig.\ref{fig:alldata_vs_Wp}(c) correspond to clean graphene ribbons with clean edges at given $B$ and $V_g$ parameters, while the nature of the ribbon edges in the experimental samples, as well as the amount of disorder, are not identified and measurements are extracted for various values of $B$ and $V_g$. The main conclusion of Fig.\ref{fig:alldata_vs_Wp}(c) is the fact that numerical data which include inelastic effects reproduce qualitatively the exponential decay of the Fano Factor observed experimentally for long interfaces. Besides, the discrepancy between numerical and experimental data for small $W_p/L_\phi$ hints at the role of disorder\,\footnote{The hypothesis of clean samples with zigzag A or B edges in Ref.\cite{Kumada2015} can be excluded as $F$ is independent of $V_g$ and $B$ in that case (in the regime $(\nu_n,\nu_p)=(2,-2)$) while fluctuations of $F$ are observed. The case of clean samples with well-defined armchair A, B, or C edges is also unlikely since in the coherent limit $W_p/L_\phi = 0$, $F=1-G/G_0$ (see Eq.\eqref{eq_noise_noprobe}) should fluctuate around $0$ (armchair A) or $0.75$ (armchair B and C) when $B$ and $V_g$ are varied.} in Ref.\cite{Kumada2015} (at the edges or along the n-p interface). The interplay between disorder and inelastic effects will be discussed in the next section. It is also noteworthy that in Ref.\cite{Kumada2015}, experimental data were analyzed as being consistent with the prediction\,\cite{Abanin2007} $F=1/4$ in the limit $W_p/L_\phi \to 0$ while our numerical data strongly differ from this prediction. The difference arises from the fact that we use a different approach than the one considered in Ref.\cite{Abanin2007}. Indeed, we solve the quantum problem upon taking into account inelastic scattering processes while $F=1/4$ was derived in Ref.\cite{Abanin2007} assuming incoherent and quasielastic mixing between the interface modes, within a semiclassical approximation.

\section{Disorder effects}
\label{sec_results_dis}

In this section, we study how the previous results are modified in the presence of disorder along the \textit{n-p} interface. To mimic disorder, we use the Anderson model \textit{i.e.} we add to the Hamiltonian $H$ given in Eq.\eqref{eq_def_H} a term
\begin{equation}
H_\mathrm{dis}=\sum_i \varepsilon_i\, c_i^\dagger c_i
\end{equation}
where $\varepsilon_i$ are random numbers uniformly distributed in the interval\footnote{We used this definition but did not investigate scaling properties with $s_f$.} $[-V_\mathrm{dis}/s_f/2,V_\mathrm{dis}/s_f/2]$. The sum over $i$ is restricted to the sites of the graphene layer which are located along the \textit{n-p} interface in a rectangle of length $L_d$ and width $W_d$ ($L_d$ and $W_d$ are defined similarly to $L_p$ and $W_p$, see Fig.\ref{fig:sys}). In the following, we take $W_d=W$ and $L_d=L_p=30$\,nm. 

\begin{figure}
    \includegraphics[clip,keepaspectratio,width=\columnwidth]{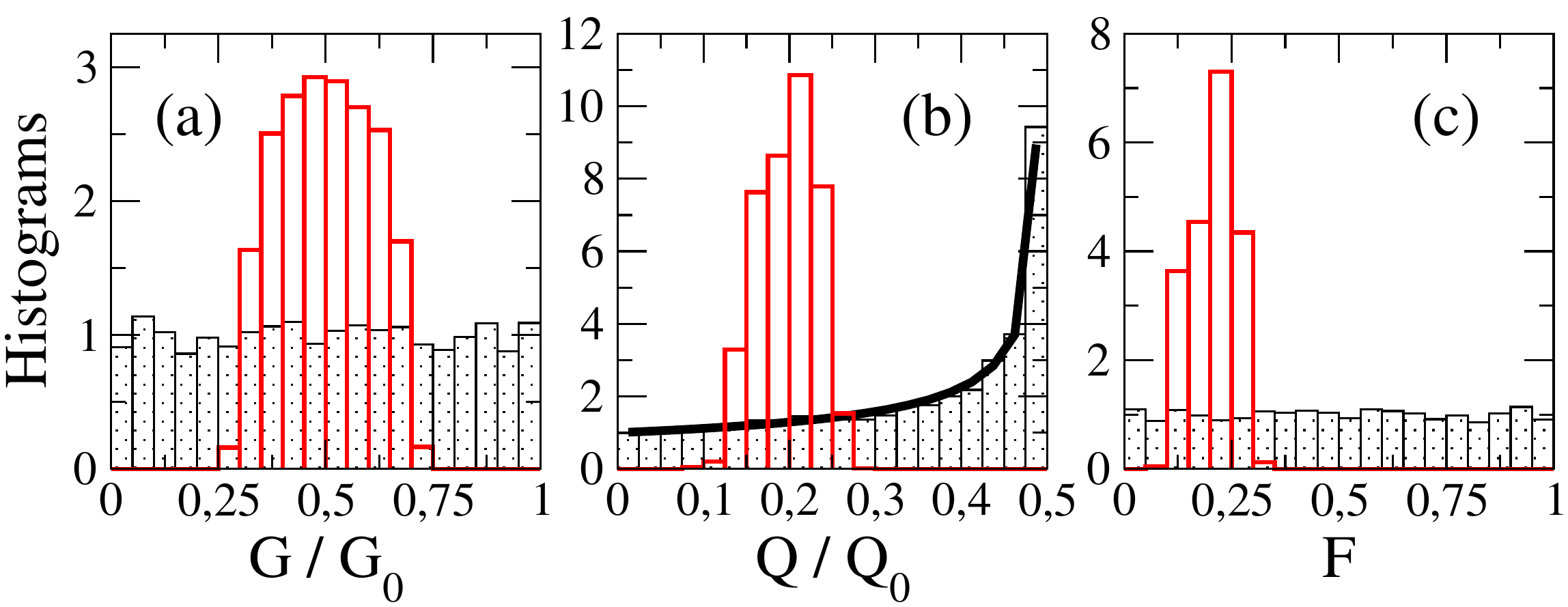}
    \caption{(Color online) Normalized histograms of $G/G_0$~(a), $Q/Q_0$ (b) and $F$ (c) values, for $W_p/L_\phi=0$ (in black, coherent case, $t_p=0$) and for $W_p/L_\phi\approx 1.7$ (in red, $t_p=0.75$\,eV, $W_p=40$\,nm). Each histogram is constructed from 4000 disorder configurations at fixed $V_{\mathrm{dis}}=2\,t_0$. The thick black line in panel (b) corresponds to $y=\int_{x-\delta/2}^{x+\delta/2}du/\sqrt{1-2u}/\delta$, $\delta$ being the histogram bin width. The nanoribbon edges are of zigzag A type with $\mathcal{N}=74$ and $\alpha_p=0.25$.}
   \label{fig:histograms} 
\end{figure}

We show in Fig.\ref{fig:histograms} the histograms of conductance, noise and Fano factor values established by considering many disorder configurations at a fixed disorder amplitude $V_\mathrm{dis}$. We compare the histograms in the coherent case (without probes) to the ones obtained when incoherent processes are included. In accordance with Ref.\cite{Chen2011}, we find that $G/G_0$ is uniformly distributed between 0 and 1 in the coherent regime and that fluctuations are much reduced in the incoherent case. Concerning noise, it is straightforward in the coherent limit that,\cite{Pedersen1998} if the conductance distribution is $p_G(G/G_0)=1$, then the noise distribution is $p_Q(Q/Q_0)=1/\sqrt{1-2Q/Q_0}$ and the Fano factor distribution is $p_F(F)=1$ in virtue of Eq.\eqref{eq_noise_noprobe}. We check in Fig.\ref{fig:histograms} it is indeed the case and we show that on the contrary the histograms are peaked in the incoherent regime. The deeper one enters the incoherent regime (\textit{i.e.} the larger $W_p/L_\phi$ is), the smaller the widths of the histograms are (data not shown).

Hereafter, we focus on the mean values $\left\langle G\right\rangle$, $\left\langle Q\right\rangle$ and
\begin{align}
\left\langle F\right\rangle & =\frac{G_0}{2Q_0}\left\langle \frac{Q}{G}\right\rangle  \\
\left\langle\left\langle F\right\rangle\right\rangle & =\frac{G_0}{2Q_0}\frac{\left\langle Q\right\rangle}{\left\langle G\right\rangle}\,.
\end{align}
Both quantities for the Fano factor have been discussed in the theoretical literature\cite{Abanin2007,Li2008,Frassdorf2016} while experimentally (in Ref.\cite{Kumada2015}), only $\left\langle F\right\rangle$ data are shown. Note that $\left\langle F\right\rangle$ and $\left\langle\left\langle F\right\rangle\right\rangle$ tend to equal each other in the incoherent limit, when fluctuations from one disorder configuration to the other are strongly suppressed. In practice, we average our data over a few dozens of disorder configurations for small $V_\mathrm{dis}$ and large $W_p/L_\phi$, and up to 8000 disorder configurations in the opposite case.

\begin{figure}
    \includegraphics[clip,keepaspectratio,width=\columnwidth]{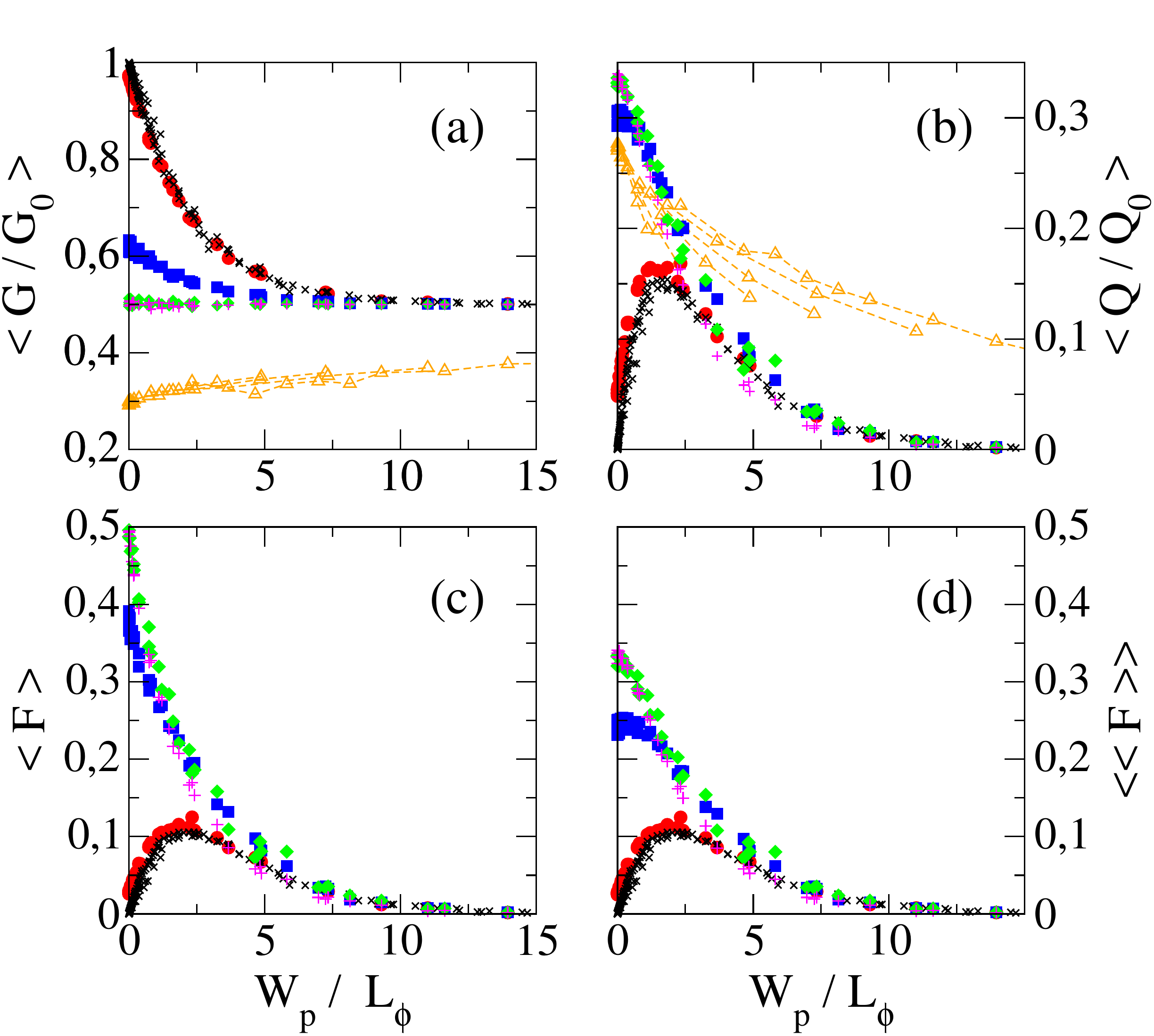}
    \caption{(Color online) Averaged $\left\langle G/G_0\right\rangle$ (a), $\left\langle Q/Q_0\right\rangle$ (b), $\left\langle F\right\rangle$ (c), and $\left\langle\left\langle F\right\rangle\right\rangle$ (d) of the disordered junction as a function of $W_p/L_\phi$. Data are plotted for different disorder amplitudes ($V_{\mathrm{dis}}=0$ ({\scriptsize{\color{black}$\times$}}), $0.1\,t_0$ ({\small{\color{red}$\bullet$}}), $0.5\,t_0$ ({\tiny{\color{blue}$\blacksquare$}}), $t_0$ ({\scriptsize{\color{green}\ding{117}}}), $2\,t_0$ ({\scriptsize{\color{magenta}$+$}}), and $6\,t_0$ ({\tiny{\color{orange}$\triangle$}})). Each curve has been obtained with $\alpha_p=0.25$, by varying $W_p$ from $10$ to $60$\,nm and $t_p$ from $0.01$ to $1.7$\,eV, upon rescaling data using the set $L_\phi(t_p)$ extracted in the clean case. The nanoribbon edges are of zigzag A type in all cases and $\mathcal{N}=74$. Error bars on the mean values are smaller than symbol size.}
   \label{fig:alldata_with_dis_zigzagA}
\end{figure}

\begin{figure}
    \includegraphics[clip,keepaspectratio,width=\columnwidth]{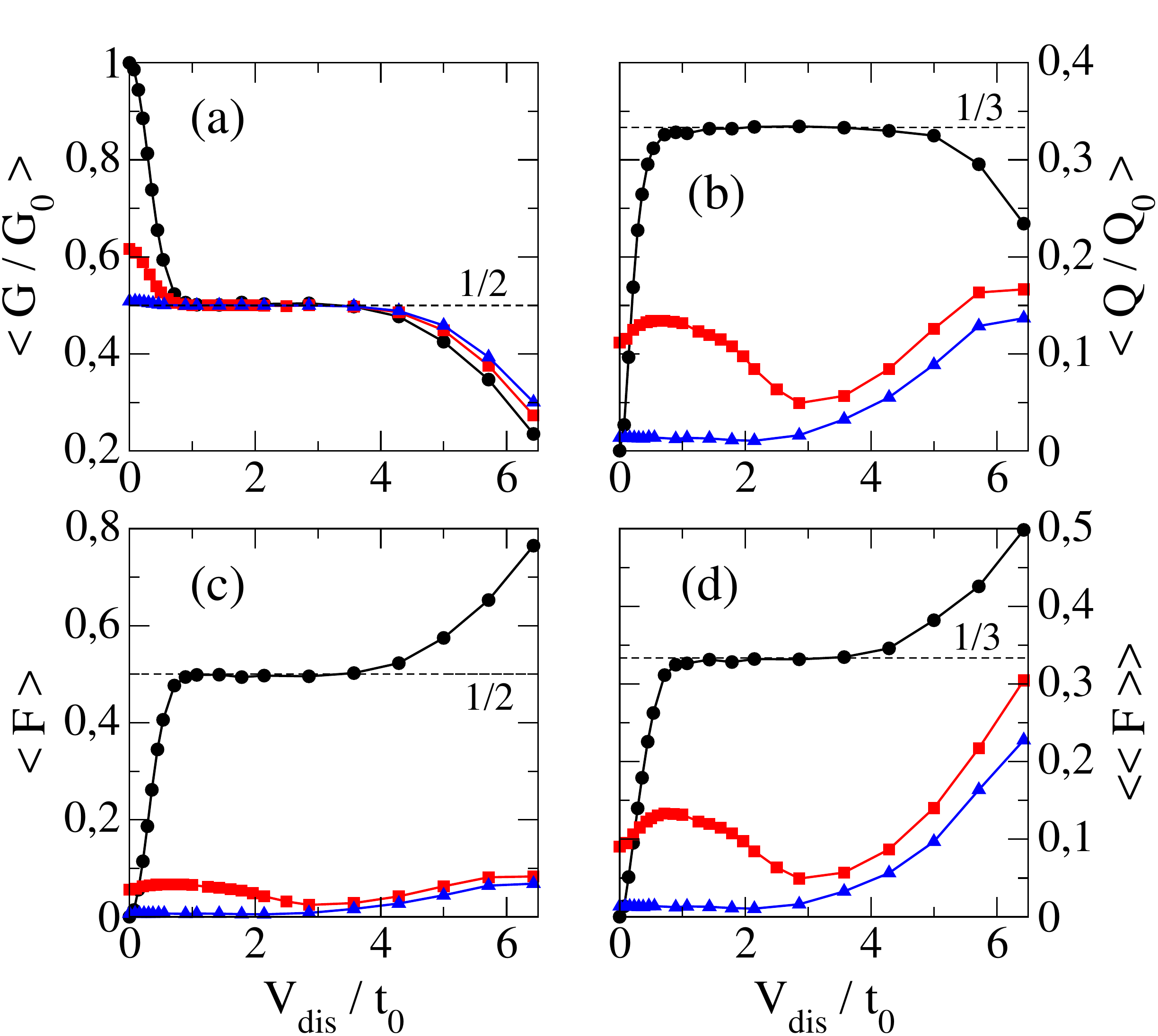}
    \caption{(Color online) Averaged $\left\langle G/G_0\right\rangle$ (a), $\left\langle Q/Q_0\right\rangle$ (b), $\left\langle F\right\rangle$ (c), and $\left\langle\left\langle F\right\rangle\right\rangle$ (d) of the disordered junction as a function of $V_{\mathrm{dis}}/t_0$. Data are plotted for different $t_p$ ($t_p=0$ ({\small{\color{black}$\bullet$}}), $1$\,eV ({\tiny{\color{red}$\blacksquare$}}), and $1.7$\,eV ({\scriptsize{\color{blue}$\blacktriangle$}})) with $\alpha_p=0.25$ and $W_p=40$\,nm \textit{i.e.} $W_p/L_\phi=0$, $3.25$ and $9.3$ respectively. The nanoribbon edges are of zigzag A type in all cases and $\mathcal{N}=74$. Error bars on the mean values are smaller than symbol size. Lines are guides to the eye.}
   \label{fig:alldata_vs_Vdis_zigzagA} 
\end{figure}

\indent In Fig.\ref{fig:alldata_with_dis_zigzagA}, we show how the curves of $\left\langle G\right\rangle$, $\left\langle Q\right\rangle$, $\left\langle F\right\rangle$, and $\left\langle\left\langle F\right\rangle\right\rangle$ versus $W_p/L_\phi$ deviate from the clean limit when the disorder amplitude is increased. For each quantity, we find that our data obtained for different $W_p$ and $t_p$ keep falling onto one single curve at finite $V_\mathrm{dis}$, when they are plotted as a function of $W_p/L_\phi$. Here we used the same set $L_\phi(t_p)$ as before \textit{i.e} the one extracted without disorder by rescaling the curves $G_\mathrm{inc}/G$ (see Sec.\,\ref{sec_scaling}). This is justified by the fact that disorder is not a source of decoherence. We note however that the scaling breaks down at large disorder amplitudes (see the orange triangles in Fig.\ref{fig:alldata_with_dis_zigzagA}(b)).
Besides, Fig.\ref{fig:alldata_with_dis_zigzagA} reveals the existence of a finite range of disorder amplitudes in which the curves are (almost) independent of $V_\mathrm{dis}$.  To study the convergence of the curves with disorder, we plot in Fig.\ref{fig:alldata_vs_Vdis_zigzagA} $\left\langle G\right\rangle$, $\left\langle Q\right\rangle$, $\left\langle F\right\rangle$, and $\left\langle\left\langle F\right\rangle\right\rangle$ as a function of $V_\mathrm{dis}$ for three particular values of $W_p/L_\phi$. In the coherent limit ($t_p=0$), the quantities are found to saturate with $V_\mathrm{dis}$ in the interval $\approx [t_0,4\,t_0]$. In this disorder range, we find plateaus at values
\begin{align}
\left\langle G\right\rangle &=G_0/2 \\
\left\langle Q\right\rangle &=Q_0/3 \\
\left\langle F\right\rangle &=1/2 \\
\left\langle\left\langle F\right\rangle\right\rangle &=1/3
\end{align}
which coincide with the analytical and numerical predictions reported in the literature for the coherent disordered \textit{n-p} junction (respectively in Refs.\cite{Long2008,Li2008,Frassdorf2016}, Ref.\cite{Li2008}, Ref.\cite{Frassdorf2016} and Refs.\cite{Li2008,Frassdorf2016}). Those values are independent of the choice of $B$, $V_g$ and $\mu$ parameters in the bipolar regime $(\nu_n,\nu_p)=(2,-2)$. We find that the conductance plateau survives in the incoherent regime while the other plateaus for $\left\langle Q\right\rangle$, $\left\langle F\right\rangle$, and $\left\langle\left\langle F\right\rangle\right\rangle$ get destroyed. New well-defined plateaus emerge at large $W_p/L_\phi$ for $V_\mathrm{dis}\lesssim 2\,t_0$. At intermediate values of $W_p/L_\phi$, there is no clear plateau (except in a very narrow range of disorder) but the variations of $\left\langle Q\right\rangle$, $\left\langle F\right\rangle$, and $\left\langle\left\langle F\right\rangle\right\rangle$  remain small as long as $V_\mathrm{dis}\lesssim 2\,t_0$.
Note that data shown in Figs.\ref{fig:alldata_with_dis_zigzagA} and \ref{fig:alldata_vs_Vdis_zigzagA} correspond to a graphene nanoribbon with zigzag A edges. When other edges are considered,\commentout{\footnote{We performed a similar study as a function of $V_\mathrm{dis}$ for armchair B edges. We might miss an (unlikely) peculiar behavior of the curves with zigzag B, armchair A or armchair C edges.}} similar curves are obtained but disorder intervals corresponding to plateaus are different.

\begin{figure}
    \includegraphics[clip,keepaspectratio,width=\columnwidth]{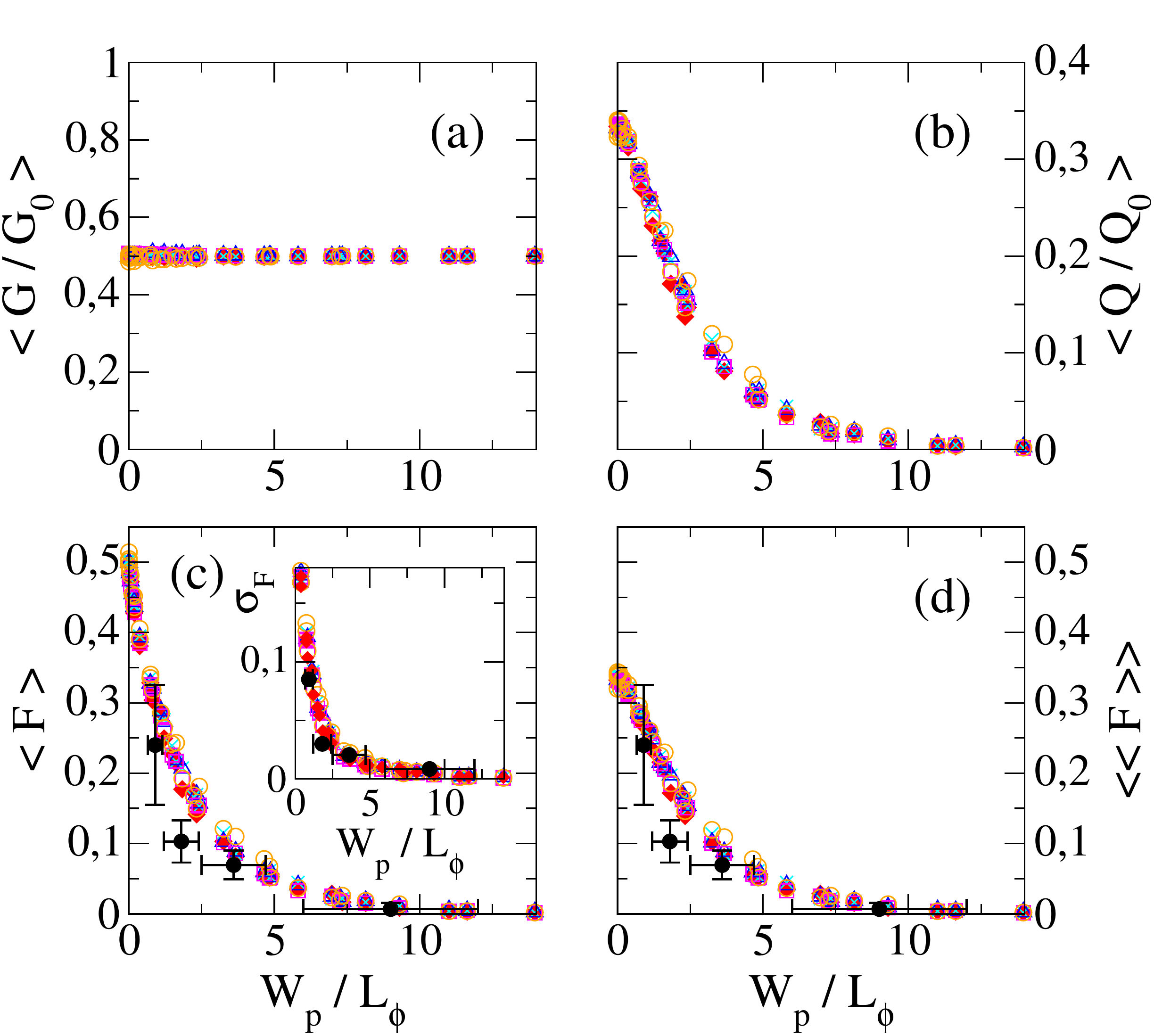}
    \caption{(Color online) Averaged $\left\langle G/G_0\right\rangle$ (a), $\left\langle Q/Q_0\right\rangle$ (b), $\left\langle F\right\rangle$ (c), and $\left\langle\left\langle F\right\rangle\right\rangle$ (d) as a function of $W_p/L_\phi$, for fixed $V_{\mathrm{dis}}=2t_0$ and different kinds of edges (zigzag A, $\mathcal{N}=74$ ({\scriptsize{\color{cyan}$\times$}}); zigzag B, $\mathcal{N}=75$ ({\scriptsize{\color{blue}$\triangle$}}); armchair A, $\mathcal{N}=65$ ({\scriptsize{\color{magenta}$\square$}}); armchair B, $\mathcal{N}=64$ ({\color{orange}{\scriptsize{\Circle}}}); and armchair C, $\mathcal{N}=66$ ({\scriptsize{\color{red}\ding{117}}})). Each curve has been obtained with $\alpha_p=0.25$, by varying $W_p$ from $10$ to $60$\,nm and $t_p$ from $0.01$ to $1.7$\,eV, upon rescaling data using the set $L_\phi(t_p)$ extracted in the clean case. Error bars on the mean values are smaller than symbol size. The black dots in panels (c) and (d) correspond to the experimental data\cite{Kumada2015} also shown in Fig.\ref{fig:alldata_vs_Wp}(c). Inset in (c): standard deviation $\sigma_F$ of the Fano factor as a function of $W_p/L_\phi$.}
   \label{fig:alldata_dis_cv} 
\end{figure}

\indent The above analysis shows us that the curves of $\left\langle G\right\rangle$, $\left\langle Q\right\rangle$, $\left\langle F\right\rangle$, and $\left\langle\left\langle F\right\rangle\right\rangle$ versus $W_p/L_\phi$ approximately converge with respect to disorder in an intermediate disorder range. In Fig.\ref{fig:alldata_dis_cv}, we fix the disorder amplitude to the value $V_\mathrm{dis}=2\,t_0$ for which convergence is reached for the five types of nanoribbon edges discussed until now. Fig.\ref{fig:alldata_dis_cv} can be seen as the disordered counterpart of Fig.\ref{fig:alldata_vs_Wp}. As expected, we find that the role of zigzag or armchair boundary conditions becomes irrelevant. Moreover, we find that experimental data\,\cite{Kumada2015} for $\left\langle F\right\rangle$ fall into the incoherent regime ($W_p/L_\phi \gtrsim 1$) where $\left\langle F\right\rangle \approx \left\langle\left\langle F\right\rangle\right\rangle$. The comparison between numerical and experimental data of the Fano factor displays a good agreement. Let us stress that black dots and vertical error bars in Figs.\ref{fig:alldata_dis_cv}(c) and (d) correspond respectively to the experimental evaluation of the mean values $\left\langle F\right\rangle$ and of the standard deviations $\sigma_F=(\left\langle F^2\right\rangle-\left\langle F\right\rangle^2)^{1/2}$ obtained in Ref.\cite{Kumada2015} by varying $B$ and $V_g$ parameters upon keeping $(\nu_n,\nu_p)=(2,-2)$. It can be argued that the variation of $B$ and $V_g$ modifies the disordered potential seen by the electronic states so that it is meaningful to compare our numerical data generated for various disorder configurations with the experimental data. In the inset of Fig.\ref{fig:alldata_dis_cv}\,(c), we also provide a comparison of the numerical and experimental\,\cite{Kumada2015} evaluations of $\sigma_F$ and find a good agreement between both within horizontal error bars.\footnote{In the limit $W_p/L_\phi \to 0$, we find $\sigma_F\to 1/\sqrt{12}$ \textit{i.e.} the standard deviation of a random variable uniformly distributed between 0 and 1.} Thus, the model including disorder and inelastic effects proves to account within data accuracy for the experimental mean values and fluctuations of the Fano factor measured in Ref.\cite{Kumada2015}.

\section{Conclusion}
\label{ccl}
Using fictitious voltage probes, we investigated the effect of inelastic scattering on electronic transport across a graphene \textit{n}-\textit{p} junction in the quantum Hall effect regime. We computed the conductance $G$, the noise $Q$, and the Fano factor $F$ of the junction at filling factors $(\nu_n,\nu_p)=(2,-2)$.
In the coherent limit, the three quantities are found to depend on the edge boundary conditions in accordance with analytical predictions reported in Ref.\cite{Tworzydlo2007}. In the opposite incoherent limit, for long interface lengths, the choice of nanoribbon edges becomes irrelevant. We provided the numerical curves describing the behavior of $G$, $Q$ and $F$ between those two limits.\\ 
\indent In the incoherent regime, we recover the experimental conductance plateau at $e^2/h$ predicted in the seminal work of Abanin and Levitov\cite{Abanin2007} and reproduced in various studies,\cite{Li2008,Long2008,Low2009,Lagasse2016,Frassdorf2016,Myoung2017} notably by a numerical approach similar to ours.\cite{Chen2011}  Our main result concerns the behavior of the noise and the Fano factor as a function of the interface length. Contrary to the conductance which saturates when the interface length is increased above the system coherence length, $Q$ and $F$ are found to be exponentially suppressed. The inclusion of disorder induces marginal corrections in the incoherent regime while it tends to suppress edge effects in the coherent regime.  
We compared our numerical results to experimental data\,\cite{Kumada2015} and found a semi-quantitative agreement without adjustable parameter, demonstrating the crucial contribution of incoherent processes to interface mode mixing.\commentout{However, we could not assert whether the experimental data for the shortest interface lengths hint at edge or disorder effects. This would require additional data in the coherent regime.} The role of disorder in graphene samples was also discussed and we reproduced with a disordered model Fano factor fluctuations measured in Ref.\cite{Kumada2015}. Our work motivates further experimental studies investigating the crossover from the quantum coherent regime to the incoherent one. This would require additional data for shorter interface lengths and cleaner samples (\textit{e.g.} in boron nitride encapsulated graphene layers).

\acknowledgments
We thank Cosimo Gorini, Norio Kumada, Patrice Roche and Xavier Waintal for interesting discussions, as well as Joseph Weston for his help in Kwant. Support from the ERC Starting Grant 679531 COHEGRAPH is acknowledged.

\bibliography{np_probes_MPRF_v3}
\end{document}